\documentclass[12pt]{elsarticle}
\usepackage{amssymb}
\usepackage{amsmath}
\usepackage{bm}
\usepackage{subfigure}
\usepackage{float}
\usepackage{natbib}
\usepackage{lscape}
\usepackage{multirow}
\usepackage{array,multirow}
\biboptions{sort&compress}
\begin{document}
\begin{frontmatter}
\title{\textbf{Unfolding problem clarification and solution validation}}
\author{N.~D.~Gagunashvili\corref{cor1}}
\ead{nikolay@hi.is}
\cortext[cor1]{Tel.: +3545254000; fax: +3545521331}
\address{University of Iceland, S\ae mundargata 2, 101 Reykjavik, Iceland }
\begin{abstract}
   The unfolding problem formulation for correcting experimental data distortions due to finite resolution and limited detector acceptance is discussed.
   A novel validation of the problem solution is proposed. Attention is drawn to fact that different unfolded distributions may satisfy the validation criteria, in which case a least informative approach using entropy is suggested. The importance of analysis of residuals is demonstrated.
\end{abstract}
\begin{keyword}
deconvolution \sep
inverse problem \sep
maximum entropy \sep
chi-square test \sep
regularisation
\PACS 02.30.Zz \sep 07.05.Kf \sep 07.05.Fb
\end{keyword}
\end{frontmatter}

\section{Introduction}
The probability density function (PDF) $P(x')$ of an experimentally
measured characteristic $x'$, differs in general, from the true physical
PDF $p(x)$.  Formally the relation
between $P(x')$\/ and $p(x)$\/ is given by equation
\begin{equation}
    P(x') \propto \int_{\Omega} p(x)A(x)R(x'|x) \,dx \;,
\label{p1_main}
\end{equation}
where $A(x)$ is the probability of recording an event with a characteristic $x$ (the acceptance);
$R(x'|x)$, is the probability of obtaining  $x'$ instead of $x$ (the experimental resolution).

If a parametric (theoretical) model $p(x,a_{1},a_{2},\ldots, a_{l})$\/ for the true
PDF is known, then the unfolding can be done by determining the parameters, e.g., by a least squares fit to the binned data \cite{zhigunov,fitgagunash,blobelpar,zechebook}.
Here, the model, that allows a description of  the true distribution by a finite
number of parameters, constitutes  \emph{a priori} information that is
required for correcting  the distortions by the experimental setup.

In contrast, model-independent unfolding, as considered in \cite{zhigunov2,
blobel,correcting,gagunashvili,schmelling,hocker,zech,agost,
gagunashvili_phystat,albert,gagunashvili_mdm}, is an underspecified   problem, and every approach to solve it requires \emph{a priori} information about the solution.  Smoothness, positiveness and shape of the solution are examples of this \emph{a priori} information.  Methods differ,
directly or indirectly, in the way \emph{a priori} information is incorporated into the result.

In this article we discuss problems of model-independent unfolding, namely  what
 the solution of the unfolding problem is and  how to validate it. 
  This is especially important because it affects our interpretation of physical data to which unfolding methods have been applied.

\section{Unfolding problem   and   solution validation}

An unfolded distribution $\hat p(x)$ can be defined as a distribution that satisfies the validation criteria.
The only objective method for validation of the solution $\hat p(x)$ is to compare the measured $P(x')$ with  Monte-Carlo reconstructed $\hat P(x')$  distributions simulated with PDF $\hat p(x)$.

In practical applications there is a sample of experimentally measured events with $P(x')$ distribution  and a sample of simulated Monte-Carlo events with $\hat P(x')$ distribution.
To  compare the two distributions  a  homogeneity   hypothesis test is used:
\begin{equation}
H_0: P(x')= \hat P(x'). \label{bla}
\end{equation}

There are several methods to test the hypothesis of homogeneity. Here we will use 
the chi-square test \cite{cramer} since it can be used for both one-dimensional and multidimensional cases.

Let us represent the distributions of events $P(x')$ and $\hat P(x')$ by two $k$-bins  histograms with $N$ and $M$ the numbers of events respectively.  The test statistic
  \vspace {-0.2cm }
\begin{equation}
X^2 =\frac{1}{MN} \sum_{i=1}^{k}{\frac{(Mn_i-Nm_i)^2}{n_i+m_i}} \label{xsquar}
\end{equation}
has approximately a $\chi^2_{k-1}$ distribution, if hypothesis $H_0$ is valid \cite{cramer}.
Here $n_i$ and $m_i$ are the numbers of events in the  $i$th  bin of the histograms.

The chi-square test $p$-value must be calculated according to the formula
\begin{equation}
p\textrm{-value}= \int_{X^2}^{+\infty}
\frac{x^{l/2-1} e^{-x/2}}{2^{l/2}\Gamma(l/2)}dx, \label {kalfa}
\end{equation}
where $l=k-1$, see Ref.~\cite{text}. Hypothesis of homogeneity must be rejected  if $p$-value is lower than a predefined significance level. Significance levels 0.1, 0.05 and  0.01  are usually used in statistical analysis.

The  validation   procedure  can include an  analysis of the residuals $r_i, i=1,...,k$ that
 is helpful for identification of the bins of the histograms responsible for a
significant overall $X^2$ value. The
 adjusted (normalized) residuals  \cite{haberman} are the most convenient for the analysis:
\begin{equation}
r_i=\frac{n_{i}-N\hat{p}_i}{\sqrt{N\hat{p}_i}\sqrt{(1-N/(N+M))(1-(n_i+m_i)/(N+M))}},
\end{equation}
 where
\begin{equation}
 \hat{p}_i= \frac{n_{i}+m_{i}}{N+M}.
\end{equation}
 If the hypotheses of  homogeneity $H_0$ are valid then the
residuals $r_i$ are approximately independent and identically distributed
 random variables  with standard normal PDF $\mathcal{N}(0,1)$.
 Analysis of the residuals increases the power of a validation procedure (see illustration Example 3 in Section 3).

Graphical methods are routinely used for analysis of residuals \cite{draper}. Graphs representing dependencies $r_i( x')$ and $ r_i (P(x'))$ can be considered. Residuals for both plots  must fluctuate near the line $r=0$ with  the same variability,  since they are approximately independent and identically distributed random variables, if hypothesis of homogeneity is valid.

  A quantile-quantile plot is used to test if residuals are random variables  with the standard normal PDF $\mathcal{N}(0,1)$  \cite{text}.

  To make a quantile-quantile plot:
  \begin{itemize}
    \item Order residuals from the smallest to the largest\\
    $r_{(1)}, r_{(2)},...,r_{(k)}$, where $r_{(i)}$ is the $i$-th smallest;
   \item Calculate  data quantiles \\
    $r_{(i)} = [(i-0.5)/k]$th data quantile;
    \item Calculate theoretical quantiles\\
    $r^{\ast}_{(i)} = [(i-0.5)/k]$th theoretical quantile,\\
    where  $r^{\ast}_{(i)}$ is the solution of the equation
    \begin{equation}
    [(i-0.5)/k]= \frac{1}{ \sqrt {2\pi } } \int_{-\infty}^{r^{\ast}_{(i)}} e^{-\frac{x^2}{2}}dx;
    \end{equation}
    \item Plot the points $(r_{(i)},r^{\ast}_{(i)})$.
  \end{itemize}
If the distribution of residuals is close to a standard normal one, the plotted points will lie close to a straight line with a slope equal to 1.

Use of the chi-square tests is inappropriate if any expected
frequency is below 1 or if the expected frequency is less than 5 in more than 20\% of bins \cite{text}. Expected frequency for the bin $i$ of the histogram representing  $P(x')$ is equal to $\hat{p}_iN$ and for the histogram representing $\hat P(x')$ is equal to $\hat{p}_iM$.

As mentioned in the introduction, solution of the main equation (\ref{p1_main}) is an underspecified problem.  It means that an infinite number of solutions  satisfying the validation criteria exists. \emph{A priori} information must be used to choose a particular solution among an infinite number of them.

If reasonable \emph{a priori} information is absent then the least informative solution of the problem can be chosen. One of the ways is  to  choose an unfolded distribution with a maximal value of entropy \cite{turchin,covan}
\begin{equation}
H(\hat p(x))=-\int \hat p(x)\ln \hat p(x),
\end{equation}
which is the solution with the lowest information content.
This approach can be helpful to avoid artifacts, such as false peaks.

An unfolded distribution can be considered as the result of measuring a probability density function by a hypothetical set-up with better resolution. A  similar interpretation of an unfolded distribution can be found in~\cite{zhigunov2,schmelling2}.

Comparison of  different unfolding
algorithms is only possible when  the same \emph{a priori} information about the solution is applied.
Blind tests of algorithms do not make any sense without a clearly defined \emph{a priori} information.
\section{Numerical examples}

\begin{figure}[H]
\begin{center}
\begin{tabular}{cc}
\hspace *{-0.8 cm}
\subfigure{\includegraphics[height=0.41\textwidth]{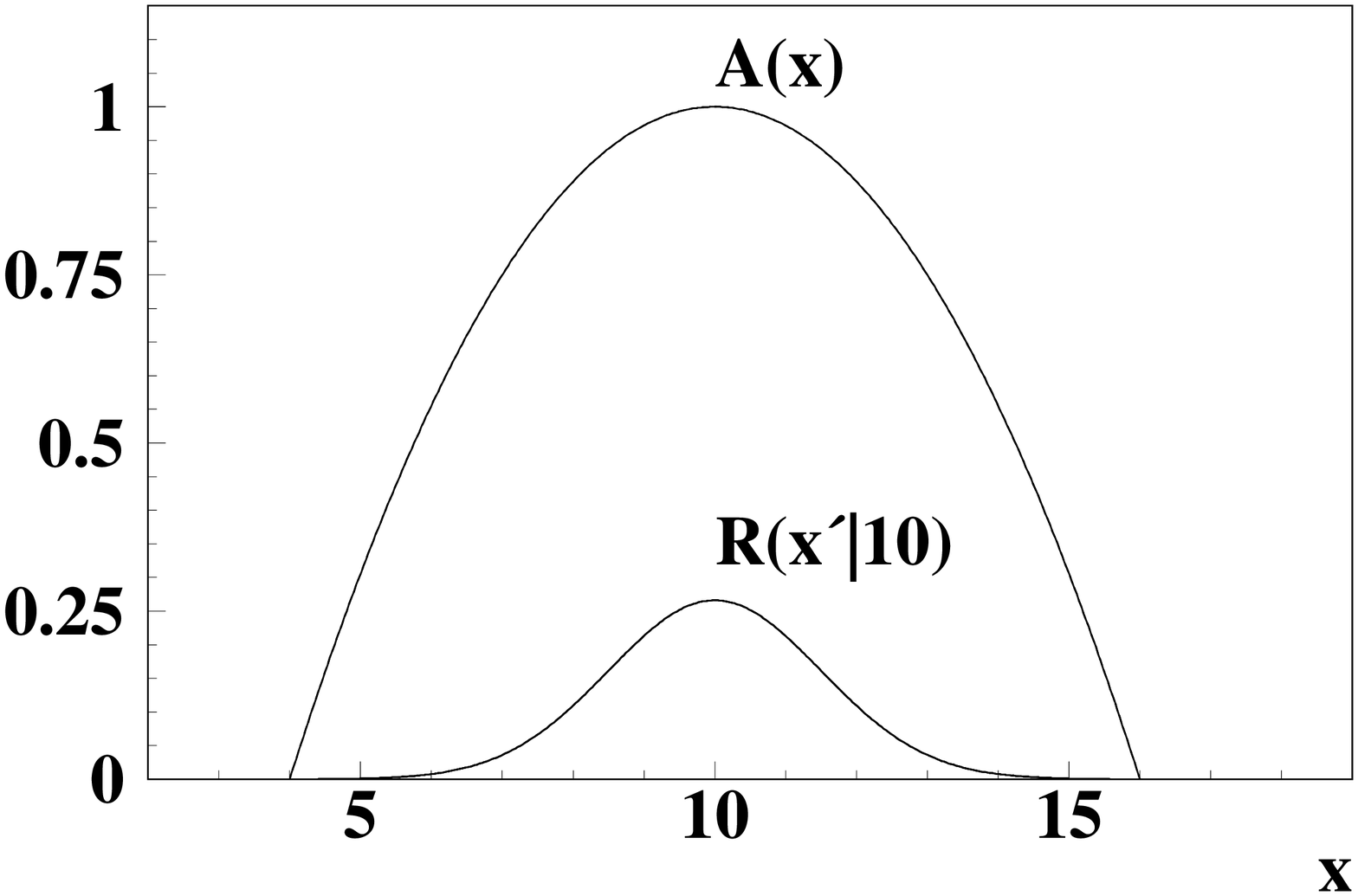}} & \hspace *{-0.8 cm}

\subfigure{\raisebox{0mm}{\includegraphics[height=0.41\textwidth]{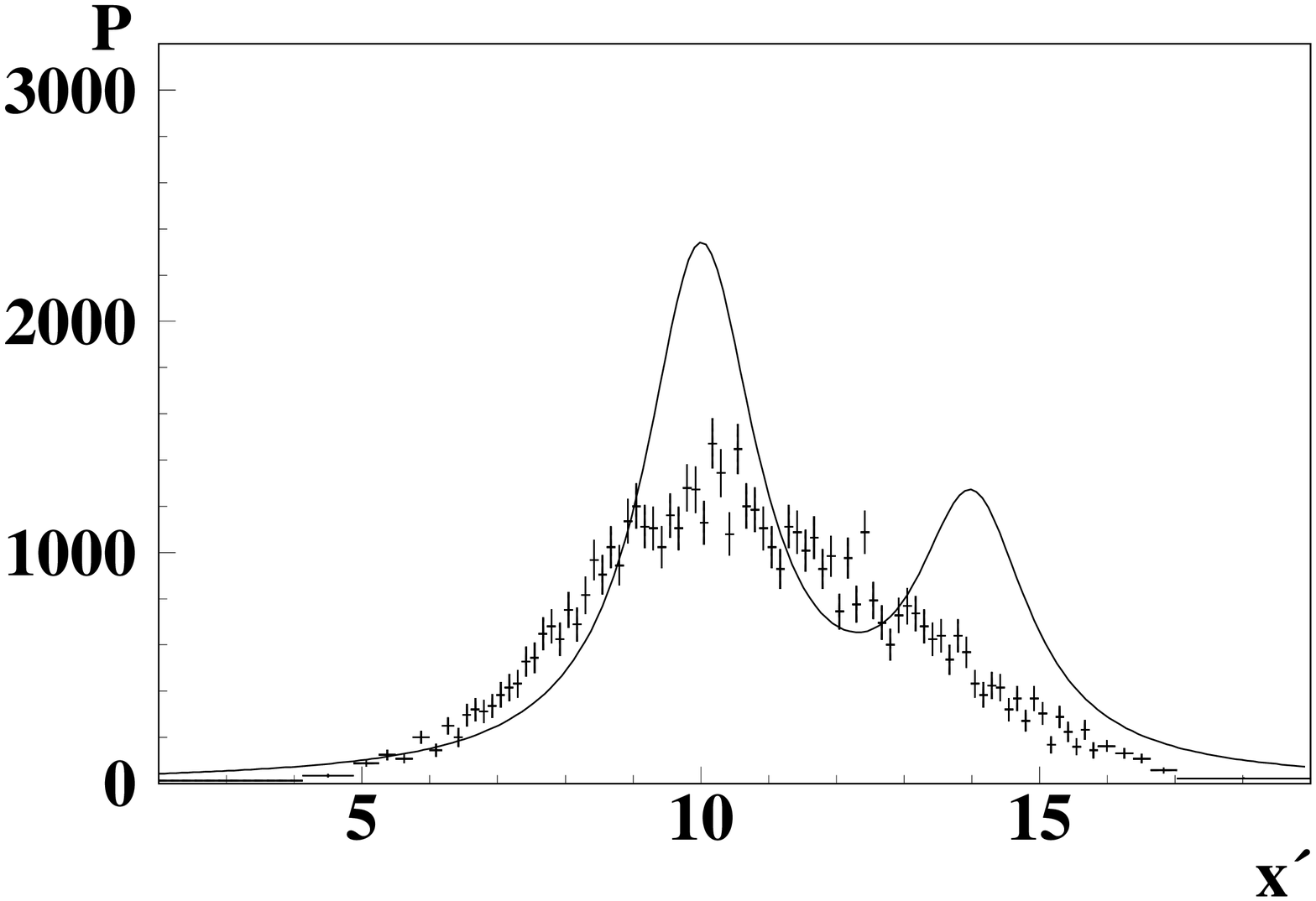}}}
\end{tabular}
\end{center}
\vspace *{-1cm}
\caption{Acceptance $A(x)$\/ and resolution function $R(x'|x)$\/
         for $x=10$ (left) and a histogram of the measured
         distribution $P(x')$\/ based on a sample of $10^4$ events generated
         for the true distribution (right). Bin contents of the
         histogram are normalised to the bin width. The true distribution $p(x)$\/ is shown  by the curve (right).}
\label{fig:true0}
\end{figure}

 Following Ref. \cite{zhigunov2} let us assume a true distribution that is described by a sum of two Breit-Wigner functions \cite{zhigunov2}
\vspace{-0.1 cm}
\begin{equation}
p(x) \propto 2\frac{1}{(x-10)^2+1}
       +     \frac{1}{(x-14)^2+1}
\label{testform}
\end{equation}
from which the experimentally measured distribution is obtained by the function
\begin{equation}
P(x') \propto \int p(x)A(x)R(x'|x)dx,
\end{equation}
with the acceptance $A(x)$:
\begin{equation}
A(x)=1-\frac{(x-10)^2}{36}
\end{equation}
and the resolution function describing Gaussian smearing (Figure~\ref{fig:true0}):
\begin{equation}
R(x'|x)=\frac{1}{\sqrt{2\pi}\sigma}\exp\left(-\frac{(x'-x)^2}{2\sigma^2}\right), \, \sigma=1.5\;.
\end{equation}
An example of the measured distribution obtained by
simulating a sample of $N=10^4$ events is also shown in Figure~\ref{fig:true0}. A histogram with a number of bins $k=89$ and an approximately equal number of events in each bin was used.
\vspace *{-1. cm}
\subsection{Example 1}

Here an application of the validation method is demonstrated for the ``ideal" case when the  unfolded distribution
$\hat p_1(x)$ coincide with the  true distribution, i.e.
\begin{equation}
\hat p_1(x)=p(x).
\end{equation}
 Ten thousand events were simulated by a random number generator with  a seed different from the one used for the measured distribution $P(x')$.

\begin{figure}[H]
\begin{center}$
\begin{array}{cc}
\includegraphics[width=9cm]{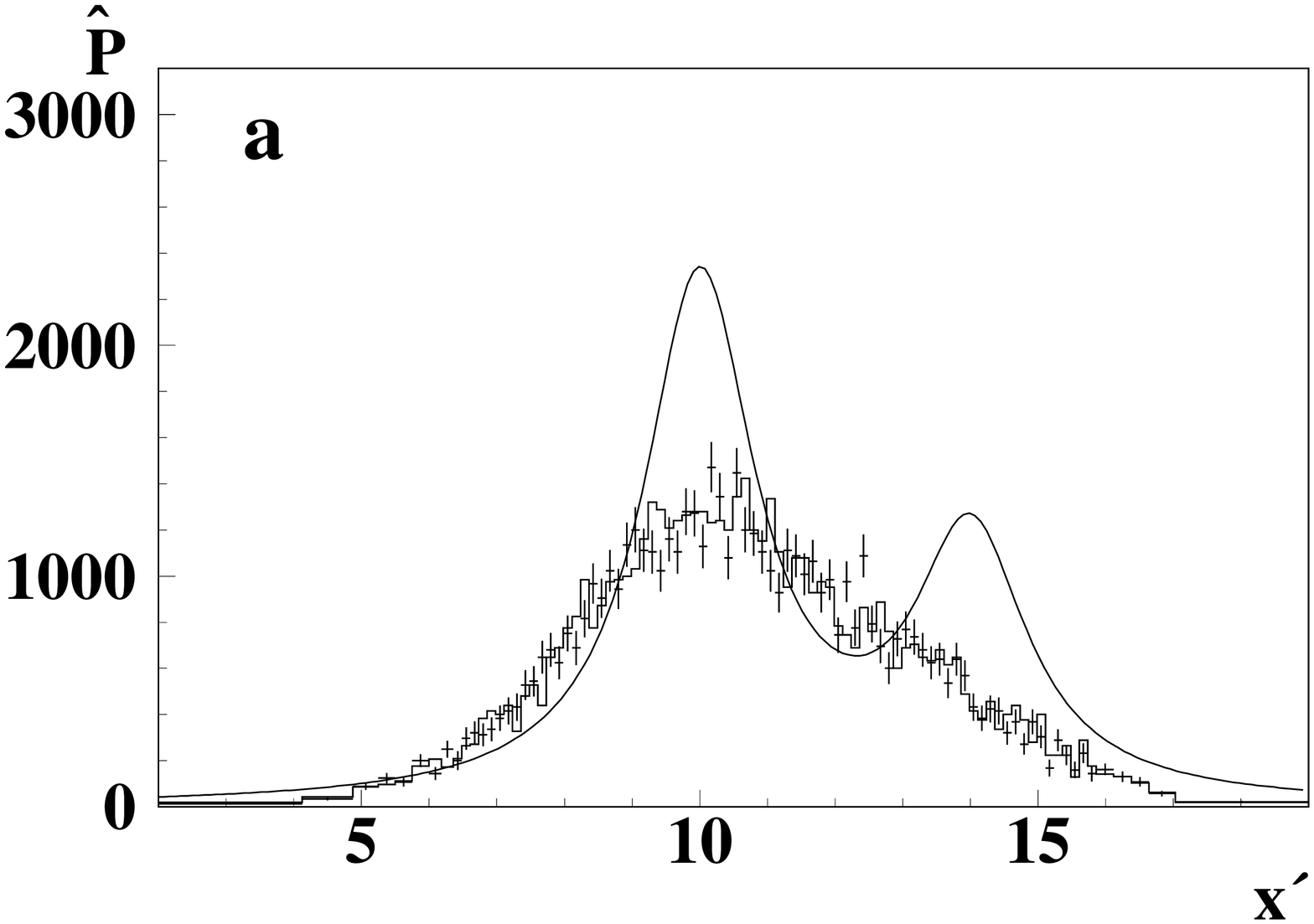} &
\vspace *{-1.1 cm} \hspace *{-0.6cm}\includegraphics[width=5.3cm]{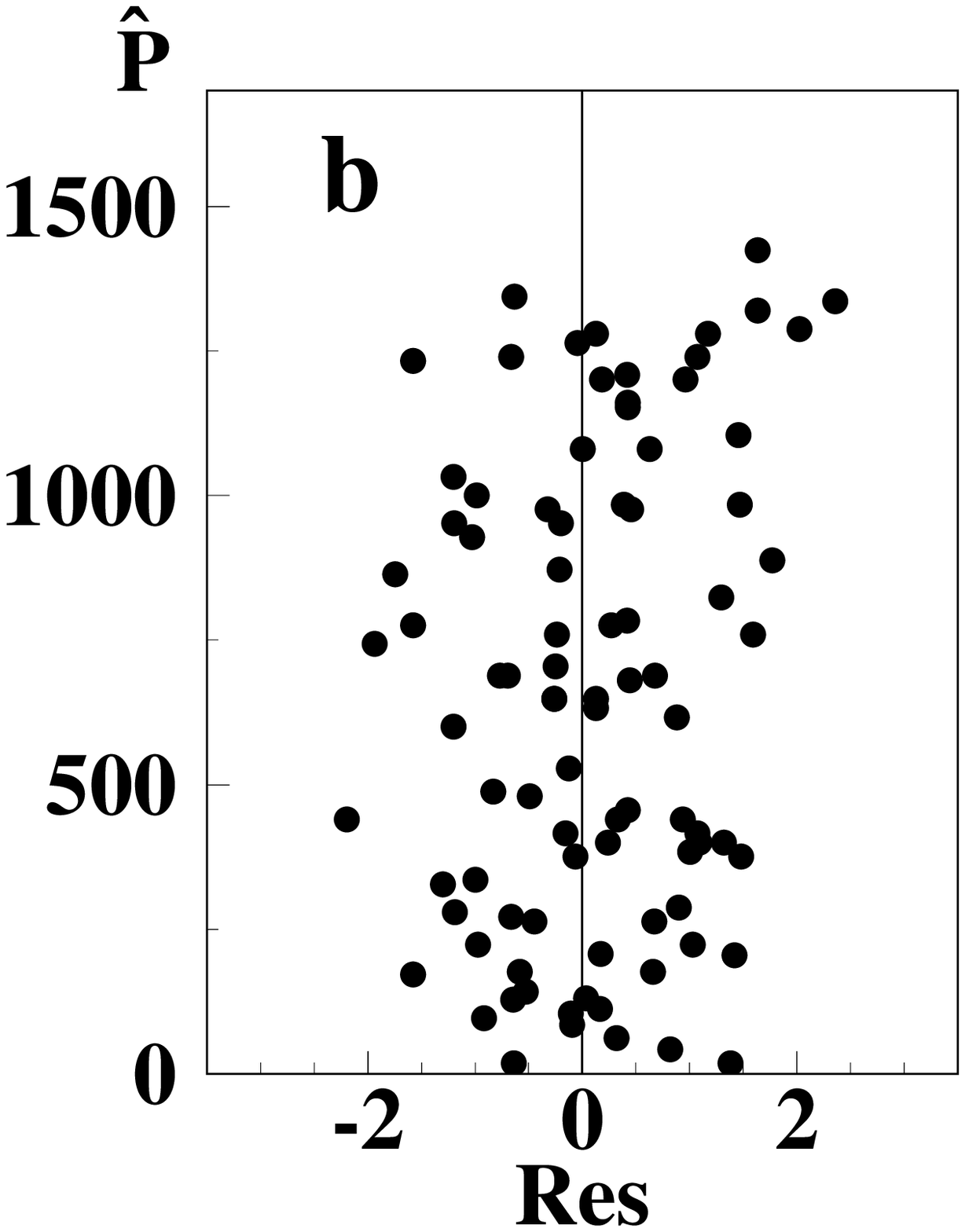} \vspace *{0.4cm}\\
\includegraphics [width=9cm]{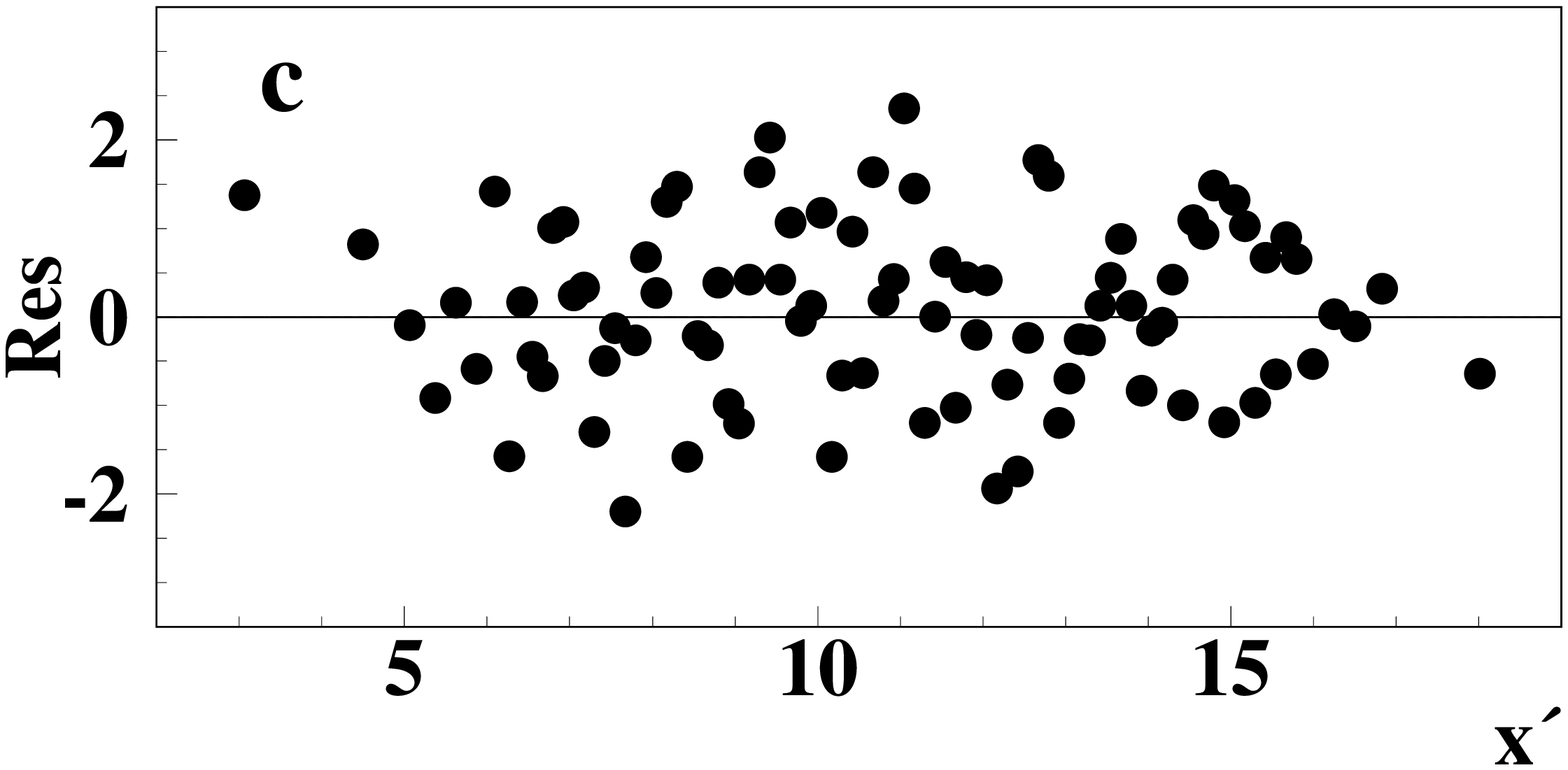}&
\vspace *{-1.1 cm} \hspace *{-0.6cm}\includegraphics[width=5.3cm]{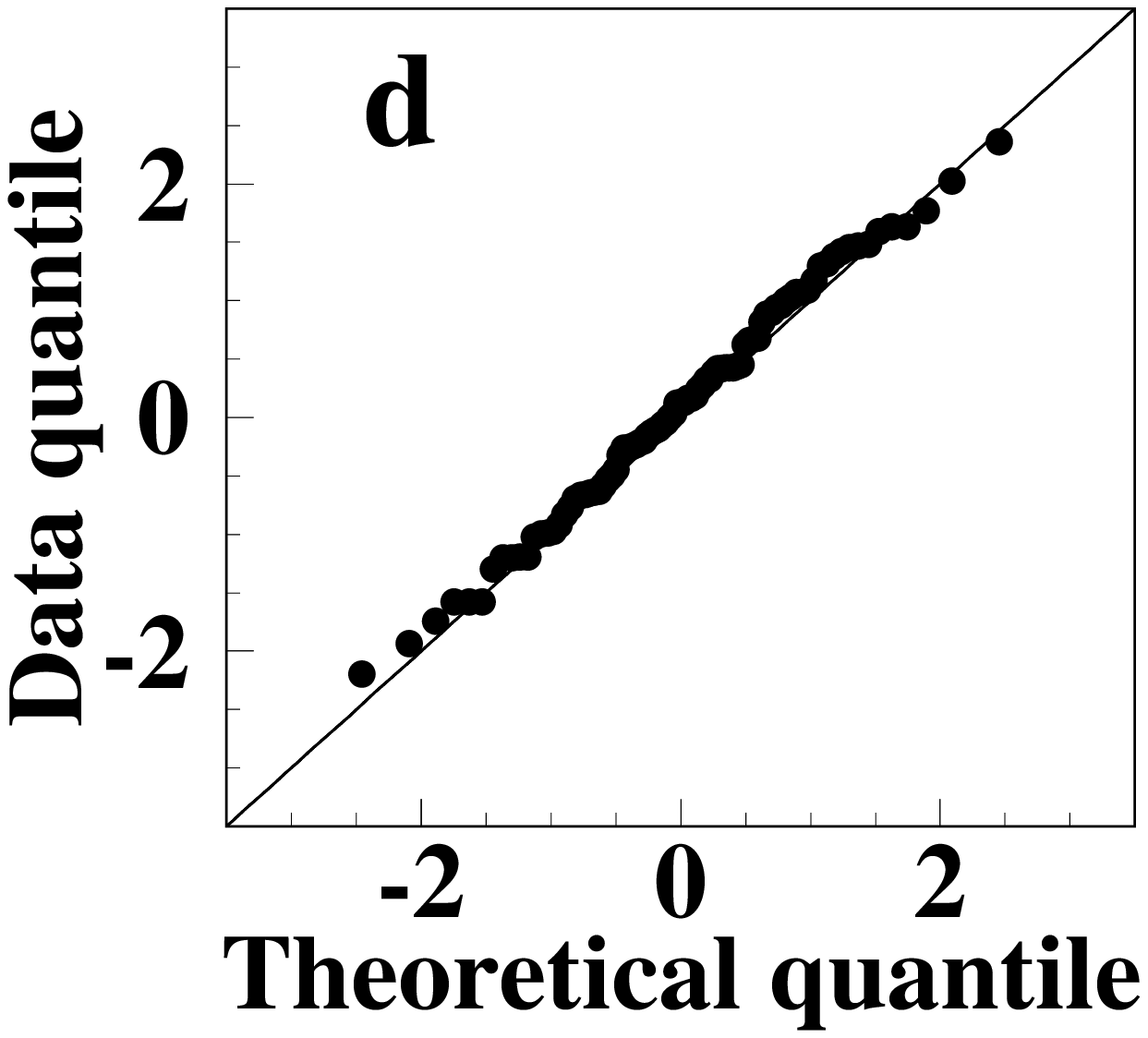}
\end{array}$
\end{center}
\vspace *{0.7cm}
\caption {
 (a)~$10^4\cdot \hat p_1(x)$ and Monte-Carlo distribution $\hat P_1(x')$ (solid lines)
 compared to the measured distribution $P(x')$ (markers with error bars);
 (b)~normalized residuals  as a function of $\hat{\bm{P}}$;
 (c)~normalized residuals as a function of $x'$;
 (d)~quantile-quantile plot for the normalized residuals.}
\label{fig:quality1}
\end{figure}

The results of unfolding validation are shown in Figure~\ref{fig:quality1}. It demonstrates
 how the Monte-Carlo reconstructed distribution $\hat P_1(x')$ (solid line), corresponding to the unfolded distribution $\hat p_1(x)$,  agrees with the measured distribution $P(x')$  (markers with error bars) (Figure~\ref{fig:quality1}a). Unfolded distribution $\hat p_1(x)$ (solid line), multiplied  by  $10^4$  for convenience,   is shown at the same plot.  Residuals and quantile-quantile plots are represented on Figures~\ref{fig:quality1}b,c,d. No structure is observed in any of the control plots. The homogeneity  test $p$-value is equal to  $0.576$, and the entropy $H(p_1(x))=2.408$.
\vspace{-0.7 cm}
\subsection{Example 2}
This example demonstrates that $\hat p_1(x)$ is not the only unfolded distribution that satisfies the validation criteria.

 Let us consider an unfolded distribution
\begin{equation}
\hat p_2(x) \propto p(x)(1+0.4\sin(5x)).
\end{equation}
 \vspace *{-2 cm}
\begin{figure}[H]
\begin{center}$
\begin{array}{cc}
\includegraphics[width=9cm]{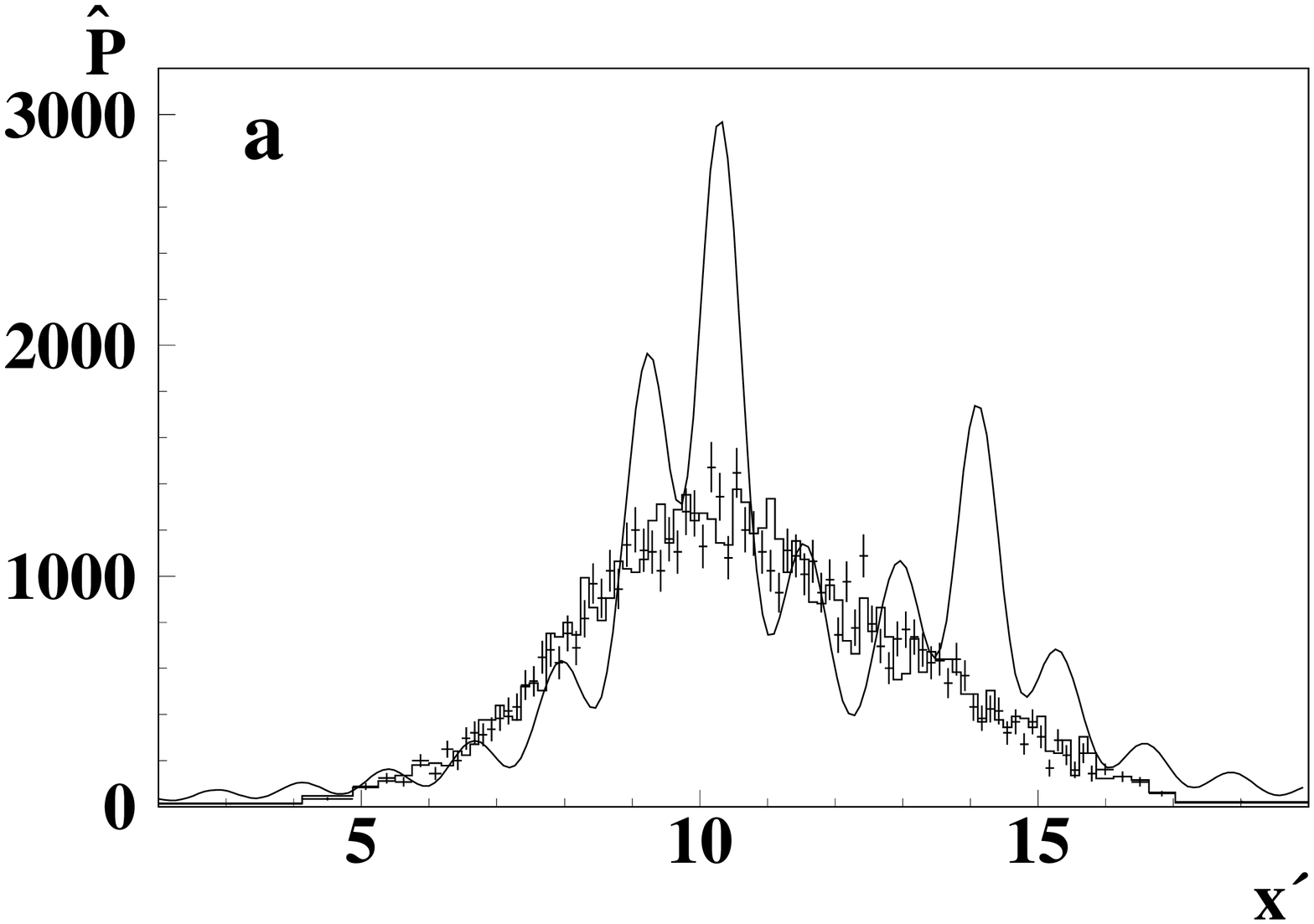} &
\vspace *{-1.1 cm} \hspace *{-0.6cm}\includegraphics[width=5.3cm]{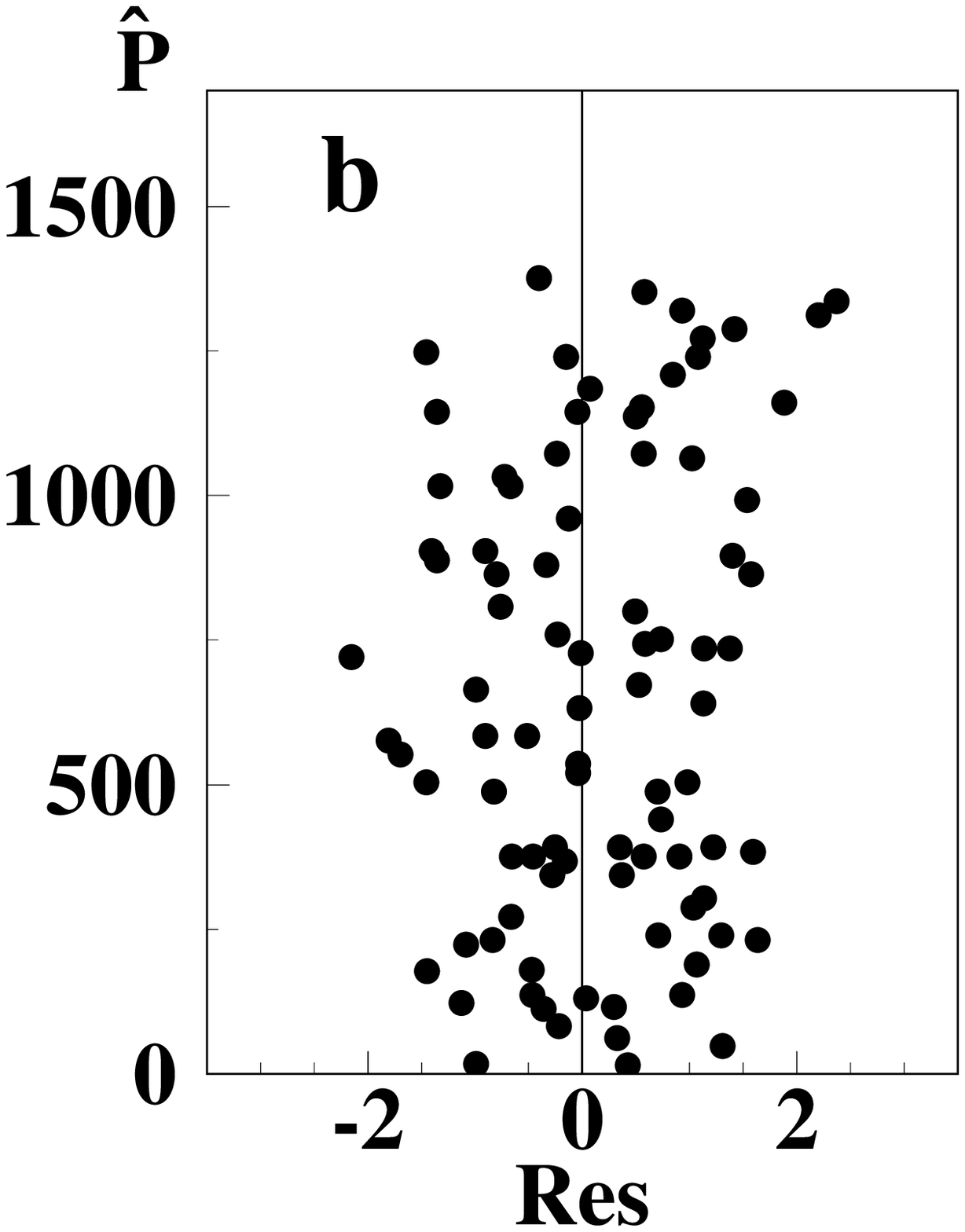} \vspace *{0.4cm}\\
\includegraphics [width=9cm]{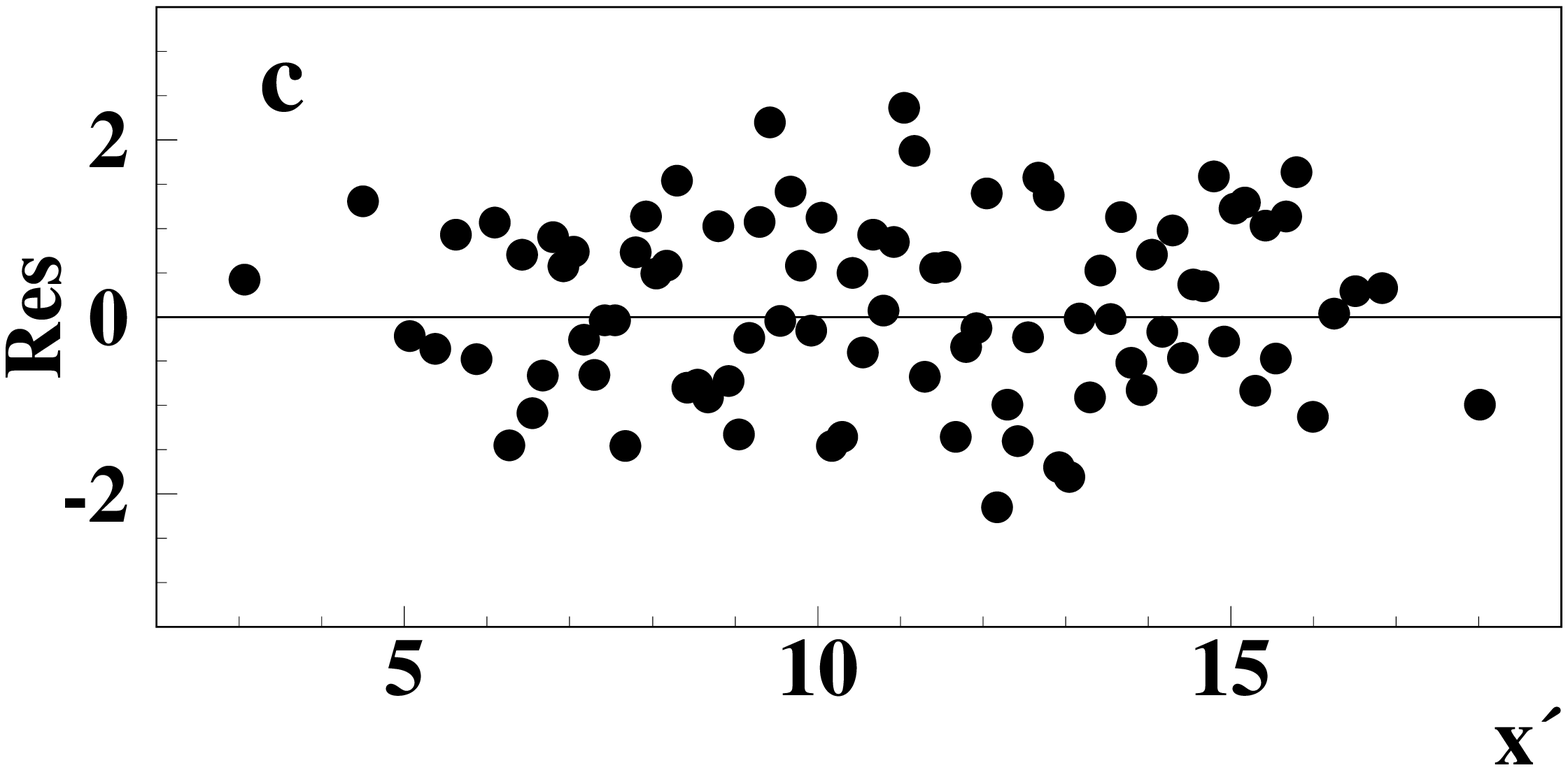}&
\vspace *{-1.1 cm} \hspace *{-0.6cm}\includegraphics[width=5.3cm]{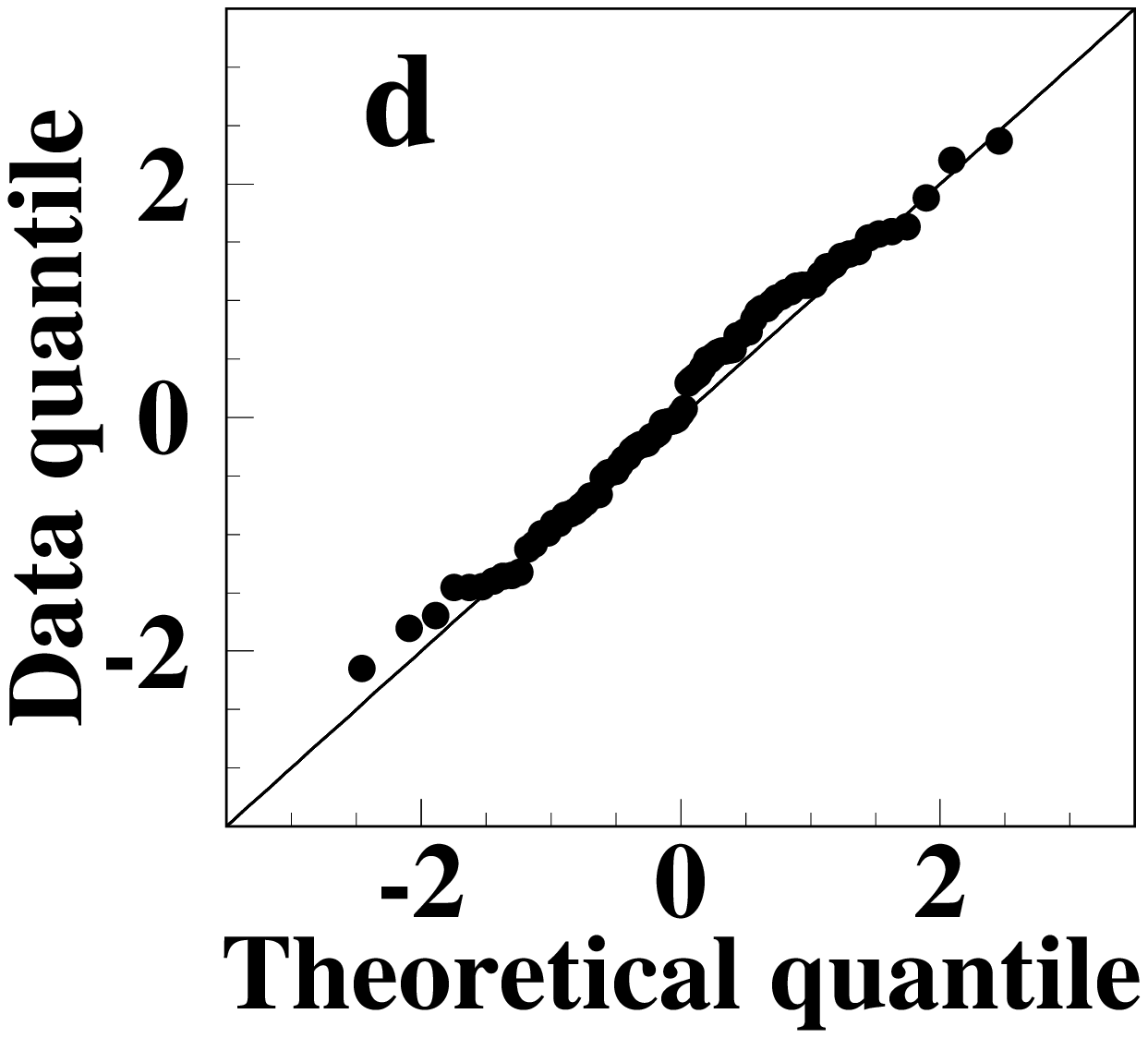}
\end{array}$
\end{center}
\vspace *{0.7cm}
\caption {
 (a)~ $10^4\cdot \hat p_2(x)$ and Monte-Carlo distribution $\hat P_2(x')$ (solid lines)
 compared to the measured distribution $P(x')$ (markers with error bars);
 (b)~normalized residuals  as a function of $\hat{\bm{P}}$;
 (c)~normalized residuals as a function of $x'$;
 (d)~quantile-quantile plot for the normalized residuals.}
\label{fig:quality2}
\end{figure}

The results of unfolding validation are shown in Figure~\ref{fig:quality2}.
It demonstrates how the Monte-Carlo reconstructed distribution $\hat P_2(x')$ (solid line), corresponding to the unfolded distribution $\hat p_2(x)$,  agrees with the measured distribution $P(x')$  (markers with error bars) (Figure~\ref{fig:quality2}a). Unfolded distribution $\hat p_2(x)$ (solid line), multiplied  by  $10^4$  for convenience,   is shown at the same plot.  Residuals and quantile-quantile plots are represented on Figures~\ref{fig:quality2}b,c,d. No structure is observed in any of the control plots.
The homogeneity  test $p$-value is equal to  $0.444$, and  entropy $H(p_2(x))=2.367$.


All these results show that the proposed PDF $\hat p_2(x)$ is also an unfolded distribution. It can be explained in terms of  signal processing.

 Acceptance  is a filter in a ``time domain" and any continuation of  the unfolded distribution $\hat p_2(x)$ outside the interval $[4; 16]$ does not  have an influence on the reconstructed distribution $\hat P_2(x)$.

 A resolution function is an acceptance in a  ``frequency domain" and is a low pass frequency filter.  A high frequency component in the unfolded distribution $\hat p_2(x)$  does not have any effect on the reconstructed distribution $\hat P_2(x)$.
Increasing the  statistics of events does not lead to rejection of the homogeneity hypothesis.

For the true distribution $ \propto p(x)(1+0.4\sin(5x))$ the least informative approach  would lead us accept   $\hat p_1(x)$ rather than  $\hat p_2(x)$ as unfolded distribution because  $H(\hat p_1(x))>H(\hat p_2(x))$.

\subsection{Example 3}
 The homogeneity hypothesis test and analyses of residuals were proposed as validation criteria. The example illustrates the importance of residual analysis.

Let us take an unfolded distribution as:
\begin{equation}
\hat p_3(x) \propto 2\frac{0.8^2}{(x-10)^2+0.8^2}
       +     \frac{0.8^2}{(x-14)^2+0.8^2}.
\label{testform2}
\end{equation}

\begin{figure}[H]
\begin{center}$
\begin{array}{cc}
\includegraphics[width=9cm]{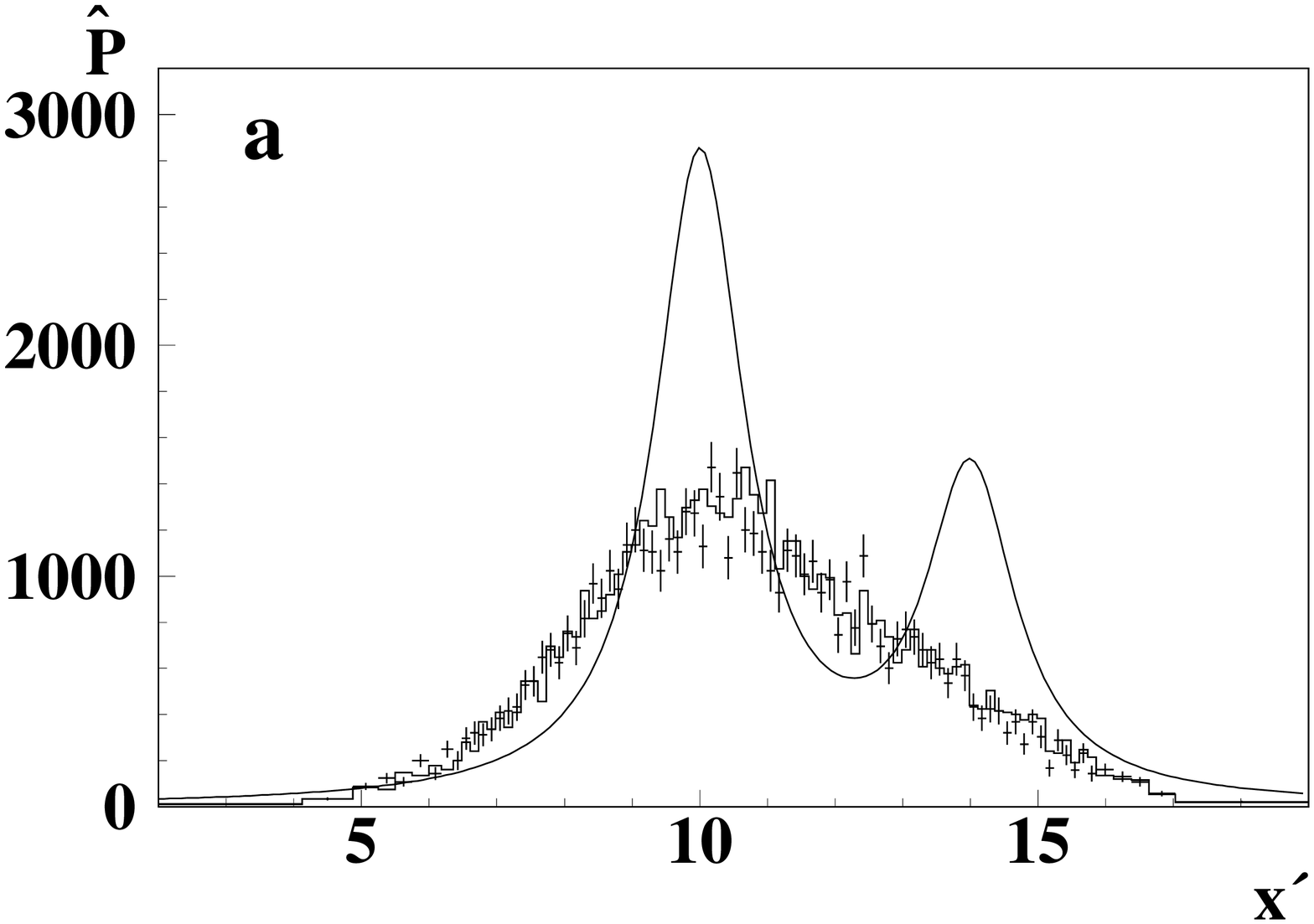} &
\vspace *{-1.1 cm} \hspace *{-0.6cm}\includegraphics[width=5.3cm]{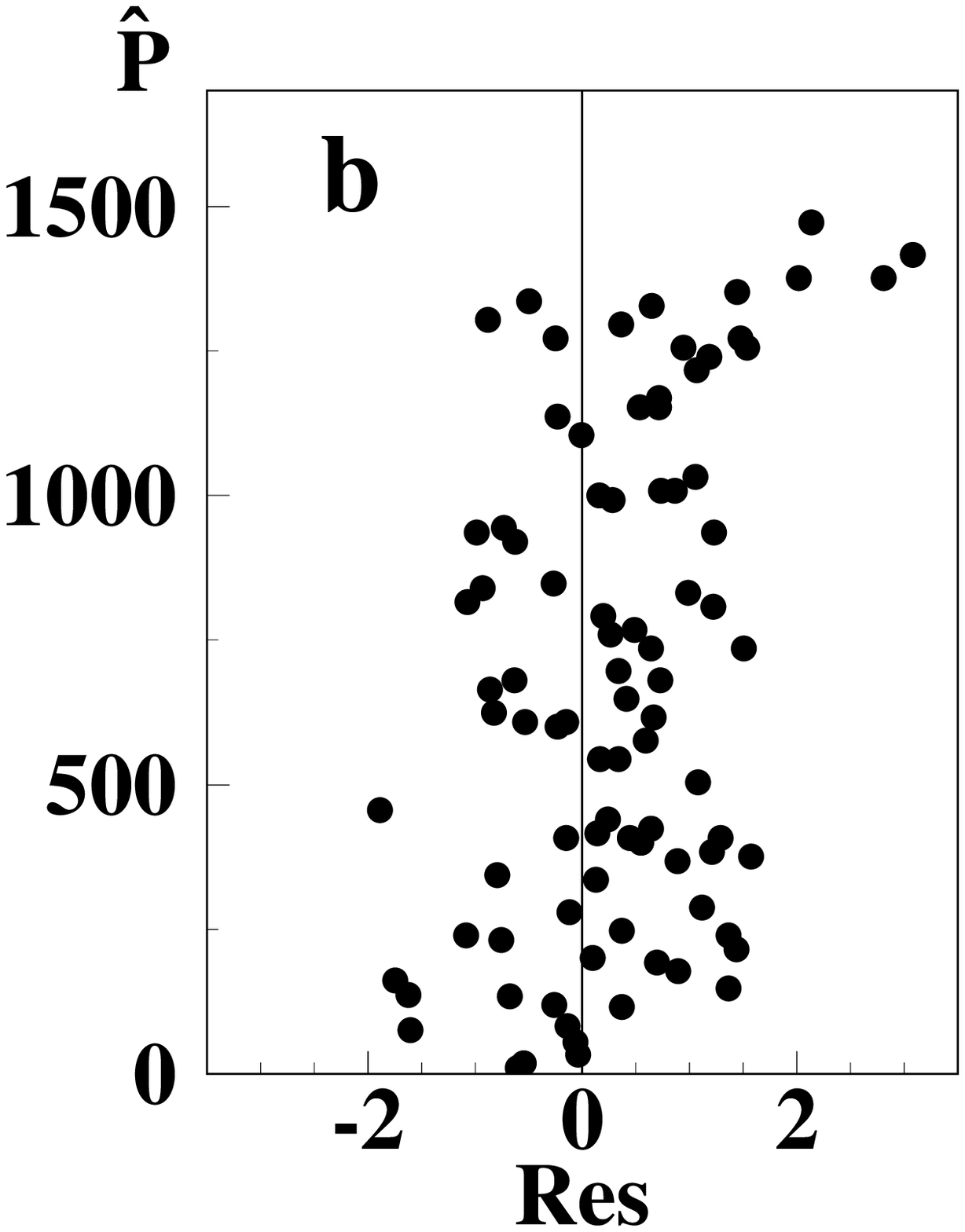} \vspace *{0.4cm}\\
\includegraphics [width=9cm]{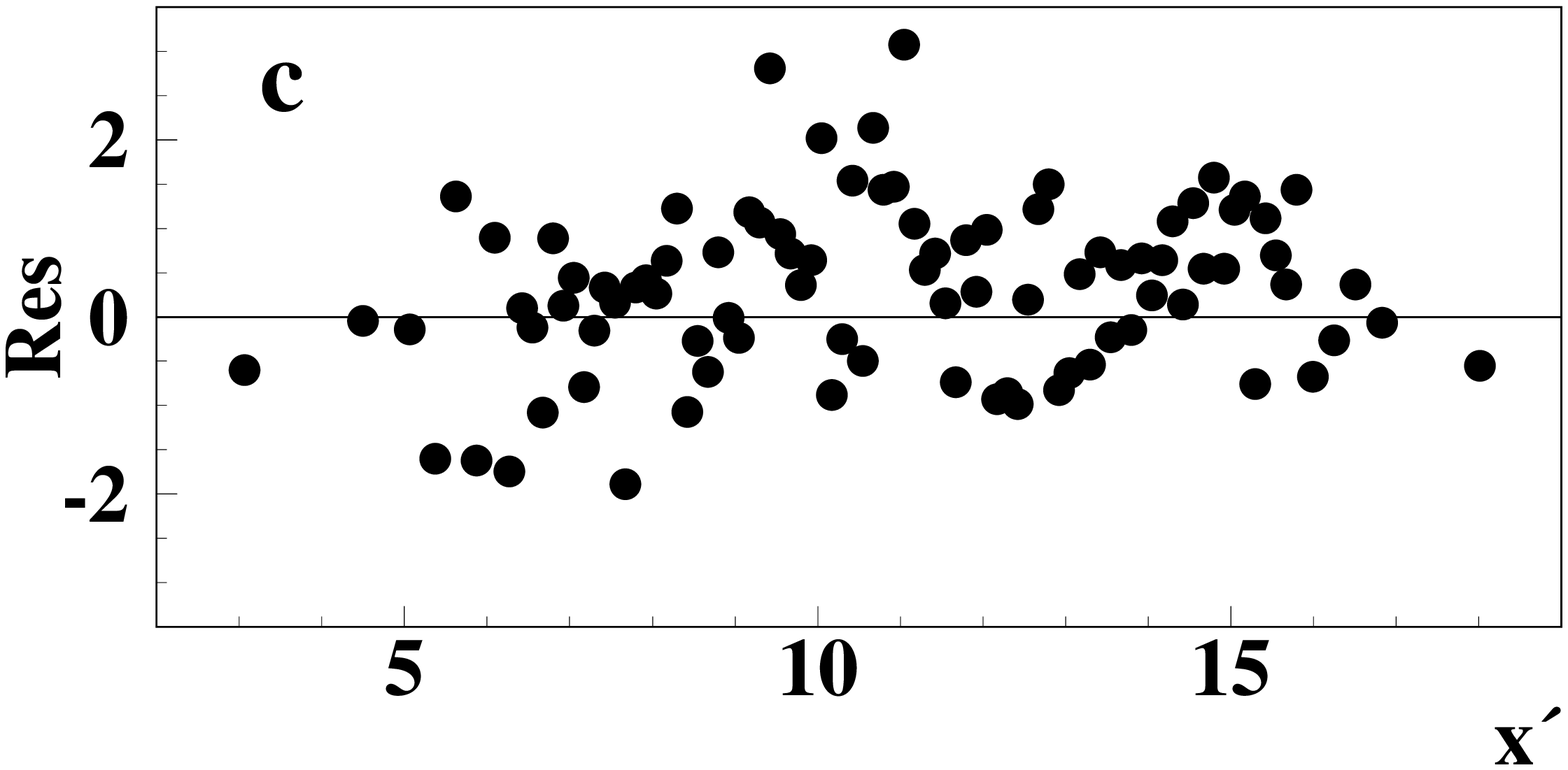}&
\vspace *{-1.1 cm} \hspace *{-0.6cm}\includegraphics[width=5.3cm]{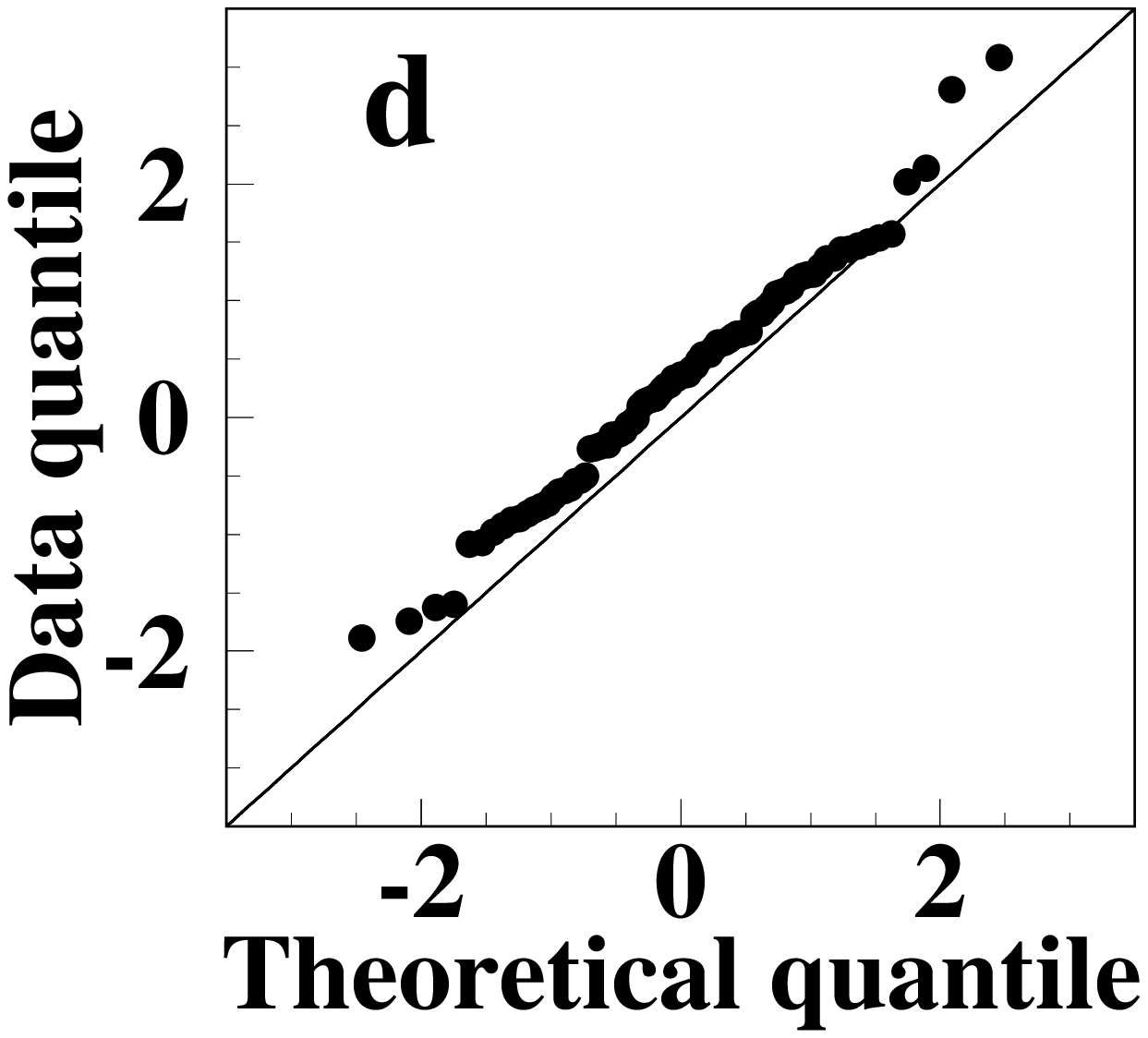}
\end{array}$
\end{center}
\vspace *{0.7cm}
\caption {
 (a)~ $10^4\cdot \hat p_3(x)$ and Monte-Carlo distribution $\hat P_3(x')$ (solid lines)
 compared to the measured distribution $P(x')$ (markers with error bars);
 (b)~normalized residuals  as a function of $\hat{\bm{P}}$;
 (c)~normalized residuals as a function of $x'$;
 (d)~quantile-quantile plot for the normalized residuals.}
\label{fig:quality3}
\end{figure}

The results of unfolding validation are shown in Figure~\ref{fig:quality3}.
It demonstrates how the Monte-Carlo reconstructed distribution $\hat P_3(x')$ (solid line), corresponding to the unfolded distribution $\hat p_3(x)$,  agrees with the measured distribution $P(x')$  (markers with error bars) (Figure~\ref{fig:quality3}a). Unfolded distribution $\hat p_3(x)$ (solid line), multiplied  by  $10^4$  for convenience,   is shown at the same plot.  Residuals and quantile-quantile plots are represented on Figures~\ref{fig:quality2}b,c,d.
 Figures~\ref{fig:quality3}b and \ref{fig:quality3}c demonstrate deviation of residual fluctuation  from the line $r=0$. Figure~\ref{fig:quality3}d demonstrates deviation of points on the quantile-quantile plot from the   straight line with slope equal to 1.
The homogeneity  test $p$-value is equal to $0.477$. However, the distribution $\hat p_3(x)$  does not satisfy validation criteria because the analysis of residuals  does not satisfy  the hypothesis of homogeneity.
\newpage
\vspace *{-2 cm}
\section{Summary and conclusions}
   A novel validation method for an unfolded distribution is proposed in this work. A solution of  the unfolding problem is defined as a distribution that satisfies validation criteria. The unfolding problem does not have a unique solution and one of the ways to define a least informative  solution to this problem is discussed here. A least informative approach to the unfolding problem can be recommended as a first step. Solid \emph{a priori} information in an unfolding procedure  can be used in a subsequent step as well as an indirect parametric fit of measured data 
    ~\cite{zhigunov,fitgagunash,blobelpar,zechebook}. In all the cases the proposed validation procedure should be applied. Numerical examples illustrate basic statements of this paper.

\section*{Acknowledgements}
\noindent
The author would like to express his great appreciation
to Michael Schmelling (Max-Planck-Institut f\"{u}r Kernphysik, Heidelberg) for encouragement and interest to this work and the author is also grateful to Sven T.~Sigurdsson (University of Iceland, Reykjavik) for careful reading of the manuscript.

\end{document}